\documentstyle[aps,prl,multicol,epsf,psfig,epsfig,verbatim]{revtex}

\begin{document}
\tightenlines
\draft
\title{Longitudinal Laser Shaping in Laser Wakefield Accelerators}

\author{Anatoly Spitkovsky$^{(1)}$ and Pisin Chen$^{(2)}$ }
\address{$^{(1)}$
Department of Physics, University of California,
Berkeley, CA 94720\\ $^{(2)}$ Stanford Linear Accelerator Center,
Stanford University, Stanford, CA 94309 }
\date{\today}

\maketitle
\begin{abstract}
We study the energetics of wake excitation during the laser-plasma 
interaction in application to laser wakefield accelerators. We find that 
both the wake amplitude and the accelerating efficiency (transformer ratio)
can be maximized by properly shaping the longitudinal profile of the driving
laser pulse. The corresponding family of laser pulse shapes is derived in the 
nonlinear regime of laser-plasma interaction. Such shapes provide 
theoretical upper limit on the magnitude of the wakefield and efficiency 
by allowing for uniform photon deceleration inside the 
laser pulse.  We also construct realistic optimal pulse shapes that 
can be produced in finite-bandwidth laser systems.
\end{abstract}

\begin{multicols}{2}

Current plasma accelerators can be broadly subdivided into two
classes -- the laser driven scheme\cite{tajima}, and the electron
beam driven scheme \cite{chen1}. 
 While the accelerating plasma waves excited in these two
schemes are very similar, the physics of wakefield
excitation is quite different between them. In the beam-driven 
Plasma Wakefield 
Accelerator (PWFA) the electron beam loses energy to the plasma through
 interaction with the induced electrostatic field, while in the 
laser-driven Laser Wakefield Accelerator (LWFA) energy loss occurs via
photon redshift or deceleration \cite{wilks}. This process is due to 
the presence of wake-induced stationary modulations in the refractive 
index of the plasma as seen in the laser comoving frame \cite{mori1}. 
With the recent advances in laser technology, the creation of 
ultrashort pulses with
virtually arbitrary temporal profiles is now possible by 
using amplitude and phase pulse shapers
\cite{shapers}.  
Since perturbations to the refractive
index of the plasma depend on pulse shape, 
different laser shapes will vary in their coupling to the plasma.
A natural question therefore arises: what laser shape
is the ``optimal'' for laser-plasma accelerators?

The number of parameters involved in selecting a particular 
 pulse shape can be overwhelming. One can characterize a shape by 
the value of its total
energy, length, maximum laser field, amplitude of plasma 
wake, etc., in addition to an infinite-dimensional space of actual shape 
functions. 
Luckily, not all of these parameters are independent or even well-defined. 
In this Letter we argue that the only two meaningful parameters that describe
a laser shape from the stand point of wake excitation are the total pulse 
energy and its depletion length.
Using these parameters 
different laser shapes can be consistently
classified and cross-compared while desired properties such as wake 
amplitude or efficiency can be optimized.

Let us consider a homogeneous unmagnetized plasma which is charge
neutral, at rest, and has initial density $n_p$ in the absence of
electromagnetic wave. Laser propagates along the $z$ axis
with initial frequency $\omega_0 \gg \omega_{p} \equiv
\sqrt{4 \pi e^2 n_p/m_e}$.
In the laser comoving frame, the
plasma response can be written in
terms of the independent dimensionless variables
$\zeta=k_{p}(z-v_g t)$ and $\tau=k_{p} c t$, where
$k_{p}$ 
is the plasma wavenumber, and  $v_g \approx -c$ is the laser group
velocity (for convenience the laser is moving in the negative $z$ 
direction). 
Introducing dimensionless normalized scalar and
vector potentials $\phi(\zeta)$ and $a(\zeta)$,
the parallel and perpendicular electric fields are
$E_\parallel =  -(m c^2 k_{p }/e) \partial \phi/
\partial \zeta$ and
${E_\perp}=-(mc/e) \partial a/\partial t=-(mc^2 k_{p
}/e) \partial a/\partial \zeta$. 
The wakefield generation equation can then be written as
\cite{sprangle2,esarey}:
\begin{equation}
{d^2 x \over d \zeta^2}= {n_e \over n_p}-1
={1\over 2}\Big({1+a^2(\zeta)\over x^2}-1\Big), \label{pot}
\end{equation}

\noindent where $n_e$ is the perturbed electron plasma density,
 $x\equiv 1+\phi$ is the modified electrostatic
potential, and $a^2(\zeta)$ is the dimensionless laser intensity envelope
averaged over fast oscillations. 
Prior to the arrival of the laser the normalized wakefield 
${\cal{E}}\equiv e E_\parallel/m_e c \omega_{p}= -dx/d\zeta$ is
zero.
A formal solution for the electric field outside the laser
can be written as the first integral of 
 (\ref{pot}): 
$[{\cal{E}}^{out}(\zeta)]^2=-{(x-1)^2/ x}+
\int_{-\infty}^{\infty} {a^2 x'/{x}^2} d\zeta$, 
which reaches a maximum value at $x=1$:
\begin{equation}
[{\cal{E}}^{out}_{max}]^2=-
\int_{-\infty}^{\infty} {a^2(\zeta) \Big ({\partial \over \partial \zeta} 
{1\over x}} \Big ) d\zeta. 
\label{eoutmax}
\end{equation}
This expression can be understood in terms of the deposition of laser 
energy into plasma. 
Due to negligible scattering, the photon number in the laser is 
essentially constant, and the laser energy loss is predominantly in the form
of frequency redshift, or photon deceleration \cite{frshift,AAC98}:

\begin{equation}
{\partial \omega \over \partial z}= -{1\over
2}{\omega_{p}^2 \over \omega}  k_{p} 
{\partial \over \partial \zeta} {n_e \over \gamma n_p} 
=-{\omega^2_{p}\over 2 \omega} 
k_{p}\Big({\partial \over \partial \zeta} {1\over x}\Big).
\label{enloss2}
\end{equation}
The energy density in the wake from (\ref{eoutmax}) can then be interpreted 
as 
the intensity-weighted integral of the photon deceleration throughout 
the pulse. Let's denote the wake-dependent part of the photon deceleration 
function as $\kappa(\zeta)\equiv x'/x^2$. It is closely related to the 
characteristic laser depletion length $l_d$, or the distance in which 
the maximally decelerated laser slice redshifts down to $\omega_p$
assuming a nonevolving wakefield. From (\ref{enloss2}),
$l_d={[({\omega_0/ \omega_{p}})^2-1] / k_{p}
\kappa_{max}}$, where $\kappa_{max}$ is the maximum of $\kappa(\zeta)$ 
inside the laser. 
The value of the peak wakefield in (\ref{eoutmax}) is bounded from above 
by the total laser energy  and the maximum photon 
deceleration:
\begin{equation}
[{\cal{E}}^{out}_{max}]^2=
\int_{-\infty}^{\infty} {a^2(\zeta) \kappa(\zeta)} d\zeta \leq
\kappa_{max} \int_{-\infty}^{\infty} {a^2(\zeta)} d\zeta .
\label{wake2}
\end{equation}

\begin{figure}
\narrowtext
\begin{center}
\epsffile{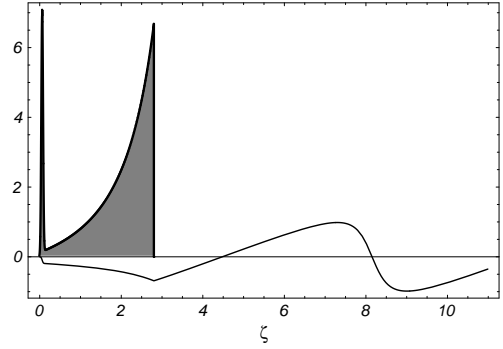}
\end{center}

\caption{General shape of the nonlinear optimal laser intensity
profile and its corresponding wakefield (arbitrary units).} \label{Shapefig}
\end{figure}

One possible optimization problem can then be formulated as follows: 
given some fixed
laser energy $\varepsilon_0$ (i.e., the integral of $a^2$), 
and the desired depletion length
(i.e., $\kappa_{max}=\kappa_0$), what laser shape would produce the largest
possible wakefield? From (\ref{wake2}) it is clear that such a shape 
should maintain a constant maximal photon deceleration throughout the pulse.
If the laser is present for $\zeta>0$, then in 
order to satisfy the boundary conditions the photon deceleration should 
rise from $0$ to value $\kappa_0$ at the very beginning of the pulse, e.g.,
like a step-function: $\kappa(\zeta)=\kappa_0 \theta(\zeta^+)$. Here,
$\zeta^+\equiv \zeta-0^+$ in order to avoid ambiguities with
the values of step-function at 0. The corresponding laser profile is
then found from the wake equation (\ref{pot}):
\begin{equation}
a_l^2({\zeta})={2 {\kappa}_0 \delta({\zeta}^+) \over
(1-{\kappa}_0 {\zeta})^4}+{4{\kappa}_0^2 \theta^2({\zeta}^+)\over
(1-{\kappa}_0 {\zeta})^5}+{1\over (1-{\kappa}_0
{\zeta})^2}-1, \label{shape}
\end{equation}
where $ \zeta \in [0,\zeta_f<1/\kappa_0]$,
 and $\delta(\zeta^+)$ is a delta-function such that 
$\int_0^{\zeta>0}\delta(y^+)dy=1$.
A schematic drawing of the optimal laser intensity variation and
its associated plasma wakefield are shown in Fig.\ {\ref{Shapefig}}.
Generally, the shape consists of a $\delta$-function at the 
front followed by a ramp in intensity which is cut off at $\zeta_f$. 
In the linear regime, when $a^2 \ll 1$,
$\kappa_0\to 0$, the ramp reduces to a triangular shape 
found in \cite{AAC98,kyoto}:
$a^2={2\kappa_0}(\delta({\zeta^+})+{\zeta})$. 
We note that (\ref{shape}) describes a family of shapes, rather than a
fixed shape. 
The actual profile of the optimal pulse depends on the deceleration 
parameter $\kappa_0$ set by the desired depletion length and the pulse
 length $\zeta_f$, which is determined from the available total energy:
\begin{equation}
\varepsilon_0 = 
2 \kappa_0 + {\zeta_f [\kappa_0^2+(1-\kappa_0 \zeta_f)^3] \over
(1-\kappa_0 \zeta_f)^4}.
\label{energy}
\end{equation}
Although the pulse length cannot exceed $\zeta_c \equiv 1/\kappa_0$,
the rise of $a^2$ towards the end of the pulse
guarantees that any finite laser energy can be accommodated for
$\zeta_f < \zeta_c$.
The two terms in (\ref{energy}) represent the energy contained in the 
$\delta$-function precursor and the main pulse. It is clear that for a fixed
total energy there exists a maximum value of $\kappa_0=\varepsilon_0/2$ 
which is achieved
when $\zeta_f \to 0$, i.e., all of the energy is concentrated in the 
$\delta$-function. This shape, which is a particular case of the 
general optimal shape (\ref{shape}), excites the largest possible wakefield
and has the smallest depletion length among all pulses of fixed energy. 
For circularly polarized pulses with cylindrical transverse 
crossection of radius $r_0$ and wavelength $\lambda$, the maximum 
achievable wake is then given by:
\begin{equation}
E_{\mathrm{max}}=6.54 E_{wb} \Big[{U_0\over 1\mathrm{J}} \Big]\Big[{\lambda\over 1\mu 
\mathrm{m}} \Big]^2
\Big[{10 \mu \mathrm{m}\over r_0} \Big]^2 
\Big[{n_p \over 10^{18}\mathrm{cm}^{-3} } \Big]^{1/2} 
\end{equation}
where $U_0$ is the total pulse energy (in Joules) and 
$E_{wb}=96 [n_p/10^{18} \mathrm{cm}^{-3}] \mathrm{GV/m}$ is the 
nonrelativistic wavebreaking field.  

\vskip -.1in
The method used for obtaining the optimal shape (\ref{shape})
is actually more general and can be used to determine laser shapes 
that 
 generate other variations in the nonlinear index of refraction.
Having a physical requirement for the refractive index 
$\eta\equiv [1-(\omega_p/\omega)^2 n_e/\gamma n_p]^{1/2}=
[1-(\omega_p/\omega)^2 1/x]^{1/2}$,
which for this case is the requirement of uniformity
of photon deceleration, provides a constraint on $x(\zeta)$, which can then be
used to find the laser shape from the wake equation. 
Alas, such a ``reverse'' solution is not always guaranteed to yield a 
physical 
(i.e., positive) $a^2(\zeta)$, so, in general, caution is advised.
\vskip .1in
While the generation of large accelerating gradients is a prerequisite for a 
successful accelerating scheme, the efficiency of acceleration should also
be considered. For an accelerating scheme that involves transfer of
 energy from 
the driver beam to the accelerating beam, efficiency is measured in terms
of the {\it transformer ratio}, or the ratio of the maximum rate of energy 
gain per particle of accelerating beam to the maximum rate
of energy loss per particle of the driving beam. In the case of laser-plasma
accelerators, where the driving and accelerating beams consist of particles 
of different species, the following kinematic definition is more useful:
\begin{equation}
R\equiv{|{\partial \gamma_a / \partial z}|_{{max}} 
\over |{\partial \gamma_d / \partial z}|_{{max}}}, \label{R1}
\end{equation}
where $\gamma_d$ and $\gamma_a$ are Lorentz factors for the driving and 
accelerating beams. In LWFA the particles in the trailing
electron bunch are accelerated via electrostatic interaction
with the wake, so ${|{\partial \gamma_a /
\partial z}|_{{max}}}  =
 |e E_\parallel^{{max}}|/m_e c^2=
 k_{p} |{\cal{E}}^{out}_{{max}}| $. For the laser propagating in plasma
$\gamma_d \approx \omega/\omega_p$, so  ${|{\partial \gamma_d /
\partial z}|}$ is the photon frequency shift given by (\ref{enloss2}). 
The transformer ratio for LWFA is then:
\begin{equation}
R^{{\mathrm LWFA}}=
{2 \omega\over \omega_{p}} 
{|{\partial x /\partial \zeta} |_{max}^{out} \over 
|{\partial ({1/ x})/ \partial \zeta} |_{max}^{in}}
 \propto {|\cal{E}|}^{out}_{max} k_{p} l_d .
 \label{R3}
\end{equation}
It follows from this definition that the transformer
ratio is not only a measure of local accelerating efficiency, but is 
also directly related to the maximum energy that can be transferred to  
the accelerating beam particle over the laser depletion length (assuming
no evolution of the wake during propagation).

As there is no absolute maximum of $R$, we can only look for a shape that
maximizes $R$ subject to constraints. For instance, among the pulses of 
fixed energy and depletion length $R$ is maximized by a pulse that produces
the largest wakefield. But this is precisely the optimal shape found above. 
A more general proof involves considering all pulses irrespective of total 
energy that create a given value of wakefield. It can then be shown 
\cite{AAC2000} that 
a pulse which has the largest depletion length among these 
must maintain a constant photon deceleration inside the pulse, which again
points to the shape (\ref{shape}). 
From (\ref{R3}) the optimal transformer ratio 
is then: 
\begin {equation}
R^{{\mathrm LWFA}}={2\omega\over \omega_{p}}
\sqrt{{1+(k_{p} L_p)^2 [1-{\kappa_0} (k_{p} L_p)]^3
\over [1- {\kappa_0} (k_{p} L_p)]^4}},
\label{TRopt} 
\end{equation}
where $L_p=\zeta_f/k_p$ is the pulse length.
In the linear regime optimal transformer ratios for both
LWFA and PWFA schemes scale identically with the pulse/beam length:
$R^{{\mathrm LWFA}}\to ({2 \omega / \omega_{p}})
\sqrt{1+(k_{p} L_{pulse})^2}, $ $R^{{\mathrm PWFA}} \to 
\sqrt{1+(k_{p} L_{beam})^2}$ \cite{chen2}. 
The LWFA scheme is intrinsically more efficient
by a factor of $2 \omega/\omega_{p}$, which is needed for viability of LWFA
since lasers are
typically ``slower'' drivers than electron beams.

\begin{figure}
\narrowtext
\unitlength = 0.0011\textwidth
\begin{center}
\epsffile{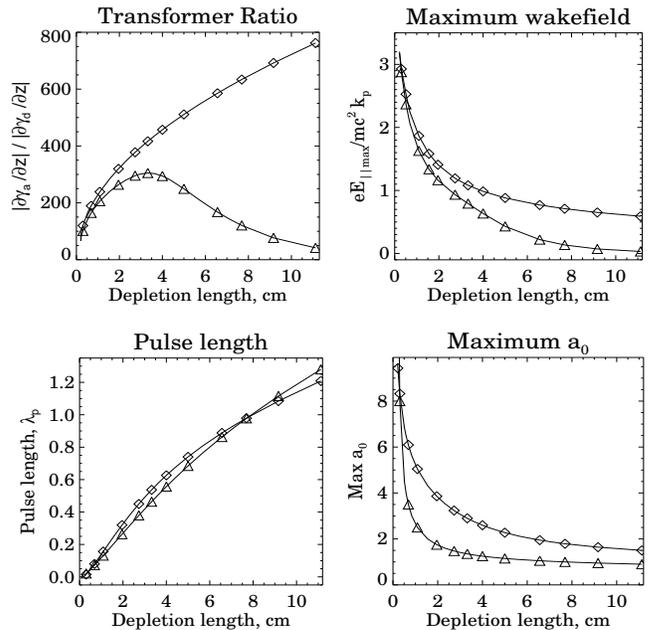}
\end{center}
\caption{
Comparison of the transformer ratio, maximum wakefield, pulse length, and
maximum normalized vector potential in shaped (diamonds) and Gaussian (triangles)
pulses of equal depletion lengths and constant pulse energy of $0.5 {\mathrm{J}}$.}
\label{TRcompar}
\end{figure}

The advantage of using the optimal pulse shape is best seen in
comparison with the unshaped (Gaussian) pulse. 
For a given
Gaussian pulse (or any other non-optimal shape) one can always
construct a corresponding optimally shaped pulse with the same laser 
energy such that the
photon deceleration across the optimal pulse equals to the peak photon
deceleration in the unshaped one 
(i.e., both pulses have equal depletion lengths). 
Unshaped pulses deplete first in
the region where photon deceleration is the largest, whereas a laser
with the optimal shape 
loses {\it all} its energy in a depletion length due to uniform 
photon deceleration, 
thus enhancing instantaneous energy deposition and wakefield.
For a
numerical example, we consider the optimal and Gaussian pulses of
total energy $0.5\mathrm{J}$, wavelength $1 \mu \mathrm{m}$ and cylindrical
radius $10 \mu \mathrm{m}$ in a plasma with $n_p=10^{18}
{\mathrm{cm}}^{-3}$.
 The transformer ratio, the maximum wakefield,
the required pulse length, and the corresponding peak $a_0$ are shown in 
Fig. \ref{TRcompar} as functions of depletion length.

From Fig.\ \ref{TRcompar} we see that the transformer ratio and the
maximum wakefield are consistently larger for shaped
 pulses. In fact, the lines for optimal pulse wakefield and transformer ratio
 represent theoretical upper limits for all pulses of given energy.
The Gaussian pulse achieves a maximum transformer ratio when its
length (measured here as FWHM) equals $1/2$ of the relativistic plasma wavelength.
The effects
of shaping are especially prominent for longer pulses, where
Gaussian pulse yields almost no wake excitation due to plasma
oscillations inside the pulse that cause part of the laser photons
to absorb energy from the wake. On the other hand, a shaped
laser postpones plasma oscillation until the end of the pulse, and 
all photons decelerate uniformly.
 For very short pulses, the differences between the 
two shapes are minimal. This
is due to the fact that very short Gaussian pulses of fixed energy 
asymptotically approach the delta-function limit of the short optimal 
shape. 

Although short pulses generally produce the largest wakefields, their efficiency
is close to minimal possible, as the depletion length decreases
faster than increase in the wake. Therefore, the choice of the 
appropriate pulse shape for LWFA stage will depend on specific
experimental conditions. If the laser-plasma interaction distance is limited
by instabilities, diffraction or dephasing, then in order to maximize 
the electron
energy gain one should try to achieve the largest accelerating gradient, which
can be accomplished with ultrashort pulses. 
If, however, the interaction length is less constrained, such as the case for 
propagation in plasma channels \cite{channels}, 
then using a finite-length shaped pulse
 will result in a greatly improved overall energy gain per stage.  
An added benefit of pulse shaping is the suppression of modulational
instability that affects unshaped pulses that are longer than 
plasma wavelength. When
all photons redshift, or ``slow down'', at the same rate,
different laser slices do not overrun each other, and the
1D laser self-modulation is suppressed. 

\begin{figure}

\begin{center}
\epsffile{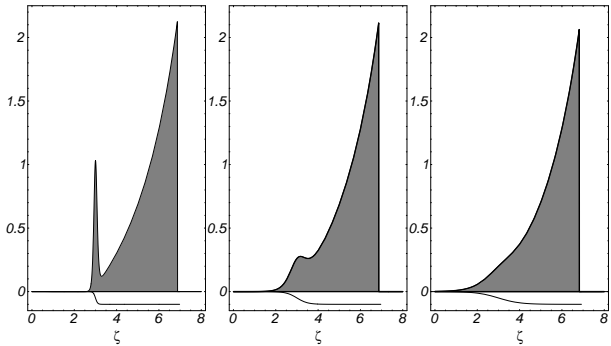}
\end{center}

\caption{Laser intensity (shaded) and the associated photon deceleration 
($-\kappa(\zeta)$)
for pulses of the same total energy and characteristic depletion length 
in the order of decreasing $\alpha$.
 }
\label{multishape}
\end{figure}

As the optimal pulse shape is
associated with a delta-function precursor, the feasibility of such
a structure may be a concern. We note that the purpose of this
precursor is to bring the photon
deceleration from zero in the quiescent plasma before the laser
to a finite value $\kappa_0$ at the beginning of the main pulse. This
 can also be achieved with a more physical prepulse, whose shape
can be found once a smooth function $\kappa(\zeta)$ is
chosen. For our example we choose $\kappa(\zeta)=\kappa_0 [1+\tanh \alpha (\zeta-\zeta_0)]/2$,
where $\alpha$ is a steepness parameter and $\zeta_0$ is an arbitrary offset.
The corresponding laser shape is then:
\begin{equation}
a^2={\kappa_0 \alpha {\mathrm sech}^2\alpha \zeta_1 \over \chi^4(\zeta_1)}+
{\kappa_0^2 [1+\tanh\alpha \zeta_1]^2 \over \chi^5(\zeta_1)}+{1\over \chi^2(\zeta_1)}-1,
\label{newlas}
\end{equation}
where $\zeta_1\equiv \zeta-\zeta_0 \leq \zeta_f$ and 
$\chi(\zeta_1)\equiv 1+(\kappa_0 / 2 \alpha)\ln{1\over 2}-\kappa_0
[\zeta_1+{1\over \alpha}\ln(\cosh\alpha \zeta_1)]/2$. As before, the pulse
length $\zeta_f$ can be found from the available pulse energy.
For a step-function photon deceleration ($\alpha \to \infty$) expression
(\ref{newlas}) asymptotes to equation (\ref{shape}). However, for finite values
of $\alpha$ the delta-function precursor spreads out and can even disappear as
shown in Fig.~{\ref{multishape}}.  The family of shapes given by (\ref{newlas})
is better suited for the finite-bandwidth laser systems that have a lower limit
on achievable feature size. The values of the maximum wakefield for pulses in
Fig.~{\ref{multishape}} are within few percent of the value for an equivalent 
delta-function optimal pulse because
the bulk of the modified laser pulse still experiences 
constant maximal photon
deceleration. The wakefield further degrades with longer rise times of
$\kappa(\zeta)$. 
It is also possible to construct optimal shapes that propagate in a 
pre-existing plasma oscillation and act as wakefield 
amplifiers~\cite{AAC2000}. Such shapes
also do not require delta-function precursors.

Several issues should be addressed before
the laser pulse shaping concept can be fully utilized.
Even without the delta-function precursor, the finite laser bandwidth 
will necessarily smooth out sharp rises and falls of the 
optimal pulse shape. Although we do not anticipate adverse effects when
the feature size is much smaller than the plasma wavelength, the 
1D self-consistent laser evolution and stability of realistic 
optimal shapes are currently under investigation. 
 Another
consideration is the influence of the laser-plasma interaction in
the transverse dimension on the evolution of the pulse. 
Many of the laser-plasma instabilities are seeded by the wakefield-induced
perturbations of the index of refraction. As we have demonstrated in this 
Letter, the nonlinear index of refraction can be effectively 
controlled through
laser shaping, thus suggesting the method of delaying the onset of 
these instabilities. Whether this approach increases the growth rates 
of other instabilities, particularly in the transverse dimension, remains to
be investigated.

We would like to thank J. Arons, A. Charman, T. Katsouleas, W. B. Mori, and 
J. Wurtele for fruitful discussions and suggestions.

\end{multicols}

\end{document}